\documentclass[preprint,aps,showpacs]{revtex4}       
\usepackage{graphicx}
\begin{document}

\title{Revised analysis of $^{40}$Ca+$^{96}$Zr fusion reactions}
\author{H. Esbensen$^1$, G. Montagnoli$^2$, and A. M. Stefanini$^3$}
\affiliation{$^1$Physics Division, Argonne National Laboratory, Argonne, Illinois 60439}
\affiliation{$^2$Dipartimento di Fisica e Astronomia, Universit\`a di Padova, and INFN, Sez. di Padova,
I-35131 Padova, Italy,}
\affiliation{$^3$INFN Laboratori Nazionale di Legnaro, I-35020 Legnaro
(Padova), Italy}
\date{\today}
\begin{abstract}
Fusion data for $^{40}$Ca+$^{96}$Zr are analyzed by coupled-channels
calculations that are based on a standard Woods-Saxon potential 
and  include couplings to multiphonon excitations and transfer channels.
The couplings to multiphonon excitations are the same as used in a
previous work. The transfer couplings are calibrated to reproduce the 
measured neutron transfer data. This type of calculation gives a poor fit
to the fusion data. 
However, by multiplying the transfer couplings with 
a $\sqrt{2}$ one obtains an excellent fit. The scaling of the transfer
strengths is supposed to simulate the combined effect of neutron and proton 
transfer, and the calculated one- and two-nucleon transfer cross sections 
are indeed in reasonable agreement with the measured cross sections.

\end{abstract}
\maketitle

\section{Introduction}

In this work we try to explain the fusion data for $^{40}$Ca+$^{96}$Zr 
by coupled-channels calculations. The data were first measured by Timmers 
et al.  \cite{timmer} and they have been 
a challenge to theory for many years \cite{back}, primarily because they 
are strongly enhanced at subbarrier energies 
(see Fig. 1 of Ref. \cite{stef4094})
compared to the data for $^{40}$Ca+$^{90}$Zr \cite{timmer} and the more 
recent data for $^{48}$Ca+$^{90,96}$Zr \cite{stefca48zr}.
It has been suggested that the enhancement is caused by the influence of 
neutron transfer reactions because the ground state Q values for neutron 
transfer are positive for this system, whereas they are negative 
for the $^{90}$Zr target (see Table 4 of Ref. \cite{timmer}). 
The expectation that the couplings to transfer channels with positive Q 
values could lead to an enhancement of subbarrier fusion was first 
proposed by Broglia et al. \cite{broglia83} in an attempt to explain 
the fusion data for the Ni+Ni isotopes \cite{becker80}.

We have previously tried to explain the fusion data for the Ca+Zr isotopes
\cite{timmer,stefca48zr} by coupled-channels calculations \cite{esbcazr}. 
We found that the couplings to multiphonon excitations play a very 
important role in explaining the enhancement of the subbarrier fusion data. 
They appeared to be sufficient to account for the $^{40}$Ca+$^{90}$Zr data
but they were clearly insufficient in explaining the $^{40}$Ca+$^{96}$Zr 
data. We tried to explain the latter data by introducing a strong coupling 
to two-neutron transfer reactions but the attempt was unsuccessful 
\cite{esbcazr}. 

In order to better understand the influence of transfer on the fusion 
of $^{40}$Ca+$^{96}$Zr, the cross sections for one- and two-neutron 
transfer reactions were measured \cite{montag2002,szil07,corradi}.
In this work we calibrate the one- and two-neutron transfer 
form factors so that the transfer data are reproduced by our 
coupled-channels calculations. We shall see that the subbarrier fusion 
is enhanced due to the couplings to the neutron transfer channels but
the enhancement is not strong enough to explain the fusion  data.
A similar conclusion was recently reached by Scamps and Hagino \cite{scamps}.
In contrast, Sargsyan et al.  \cite{sargsyan} have claimed that the fusion 
data can be explained by considering the change in the deformation of the 
reacting nuclei after the two-neutron transfer has taken place.

We mentioned in our earlier work \cite{esbcazr} that the effective 
ground state Q values for one- and two-proton transfer reactions 
are positive in $^{40}$Ca+$^{96}$Zr collisions. The couplings to 
these reaction channels could therefore also lead to an enhancement 
of the subbarrier fusion cross sections. 
A rough estimate of the combined effect of the couplings to the 
neutron and proton transfer channels is to multiply the neutron 
transfer couplings with a factor of $\sqrt{2}$. 
We shall see that this simple estimate gives an excellent account
of the fusion data. It also gives a fairly reasonable account of 
the measured one- and two-nucleon transfer cross sections,
although a detailed comprehension of these data is outside 
the purpose of the present work.

The manuscript is organized as follows. The results of a previous analysis
of the $^{40}$Ca+$^{96}$Zr fusion data, and the challenges that remain,
are summarized in the next section. 
The model that is used to describe the influence of inelastic excitations 
and nucleon transfer reactions on heavy-ion fusion cross sections is 
reviewed in Sec. III. The sensitivity to multiphonon excitations and 
multinucleon transfer reactions is investigated in Sec. IV, and the 
conclusions are presented in Sec. V.

\section{Summary of previous results}

The formalism for the coupled-channels calculations we perform is
described in detail, for example, in Sec. III of Ref. \cite{back}. 
The formalism was applied in Ref. \cite{esbcazr} to analyze the data for 
the fusion of the Ca+Zr isotopes \cite{timmer,stefca48zr}. Most of the data 
were explained fairly well by considering multiphonon excitations with up 
to three-phonon excitations, and a relatively modest influence of nucleon
transfer reactions. One exception was the fusion of $^{40}$Ca+$^{96}$Zr
which could not be reproduced at subbarrier energies, even when a very
strong pair-transfer coupling was applied. 

Another interesting feature of the analysis for the Ca+Zr fusion data
is that most of the data were best reproduced by applying the so-called
M3Y+repulsion, double-folding potential \cite{esbcazr}.
Having determined the densities of the reacting nuclei it is possible
to predict the ion-ion potential.
It turned out that the ion-ion potential that could be predicted for 
$^{40}$Ca+$^{96}$Zr produced a Coulomb barrier that was much too high.
The data were much better described by applying 
an ordinary Woods-Saxon (WS) potential, with a 1.5 MeV lower Coulomb 
barrier.  Such a potential will therefore be applied in the following. 

\begin{figure}
\includegraphics[width = 8cm]{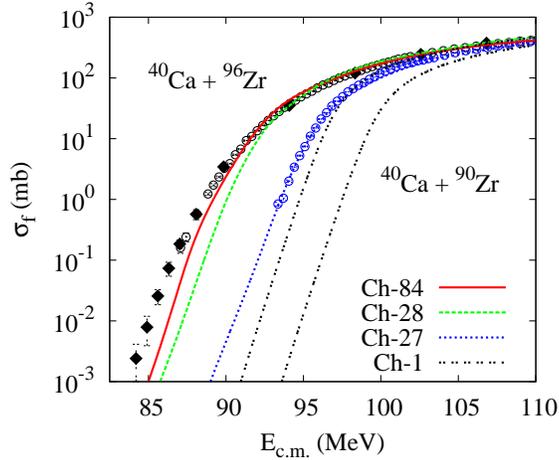}
\caption{ 
Fusion cross sections for $^{40}$Ca+$^{90,96 }$Zr \cite{timmer}.
The solid diamonds are the new data for $^{40}$Ca+$^{96}$Zr \cite{stef4096}.
The curves are based on standard WS potentials.
The Ch-1 calculations show the no-coupling limit for the two systems. 
The Ch-27 and Ch-28 calculations include couplings of up to three-phonon 
excitations. The Ch-84 calculation for $^{40}$Ca+$^{96}$Zr includes in 
addition couplings to one- and two-neutron transfer channels.}
\label{40zrffws}
\end{figure}

The $^{40}$Ca+$^{90,96}$Zr fusion data by Timmers et al. \cite{timmer}
are compared in Fig.  \ref{40zrffws} to coupled-channels calculations
that are based on standard WS potentials of the proximity type
described by Eqs. (III.40-41) and (III.44-45) in Ref. \cite{winther}.  
The parameters of the calculations and the channels that were included
are discussed in detail in Ref. \cite{esbcazr}.
The data for $^{40}$Ca+$^{96}$Zr were recently supplemented with new
measurements \cite{stef4096}. They are shown in Fig.  \ref{40zrffws} by
the solid black diamonds and reach cross sections as small as 2.4 $\mu$b.

The two Ch-1 calculations farthest to the right in Fig. \ref{40zrffws}
show the no-coupling limits for the two systems.
The Ch-27 calculation for $^{40}$Ca+$^{90}$Zr (blue dashed curve) includes
couplings of up to three-phonon excitations, with a total of 27 channels.
The Ch-28 calculation for $^{40}$Ca+$^{96}$Zr is similar and has 28
channels (the green dashed curve).
It is seen that the Ch-27 calculation reproduces the data for the $^{90}$Zr
target quite well, whereas the Ch-28 calculation underpredicts the data
for the $^{96}$Zr target at subbarrier energies.
The qualitative reason for the difference in these results is
(as mentioned in the introduction) that the ground state Q values for 
neutron transfer reactions are positive for the $^{96}$Zr target, 
whereas they are negative for the $^{90}$Zr target.
The influence of transfer is therefore expected to play a major in the 
fusion of $^{40}$Ca+$^{96}$Zr, and only a minor role in the fusion of
$^{40}$Ca+$^{90}$Zr.

In an attempt to explain the $^{40}$Ca+$^{96}$Zr fusion data, the combined
effect of multiphonon excitations and couplings to one- and two-neutron 
transfer reactions were included in Ref. \cite{esbcazr} in Ch-84 calculations.
The result is shown by the solid red curve in Fig. \ref{40zrffws}.
The WS potential and the strength of the pair-transfer coupling were
adjusted to optimized the fit to the high energy fusion data. 
This procedure failed and resulted in the discrepancy with the low-energy 
data that can be seen in Fig. 1. In the following sections we investigate 
what could be the reason for the failure.

\section{Coupled-channels calculations}

The WS potential that was used previously for $^{40}$Ca+$^{96}$Zr 
\cite{esbcazr} is adopted here for simplicity.  It has the parameters 
$R$ = 9.599 fm, $V_0$ = -73.98  MeV, and diffuseness $a$ = 0.673 fm. 
The total potential for angular momentum $L$ = 0 has a Coulomb barrier height 
of 96.62 MeV and a pocket of 73.62 MeV. The pocket is safely above the ground 
state energy of the $^{136}$Nd compound nucleus which is $E_{CN}$ = 41.089 MeV.

We also use the same one-, two-, and three-phonon excitations that were 
used in Ref. \cite{esbcazr}. The calculation with one- and two-phonon
excitations has 18 channels and is referred to as the Ch-18 calculation.
The calculation with up to three-phonon excitations has 28 channels and 
is called the Ch-28 calculation.

The influence of transfer is modeled as first described in Ref. \cite{landow}.
The model has since been used in several publications, including the study of
multineutron transfer reactions in $^{58}$Ni+$^{124}$Sn collisions 
\cite{esbnisn} and our recent work on the fusion of Ca+Zr isotopes 
\cite{esbcazr}. Since the calculations failed to reproduce the fusion data
for $^{40}$Ca+$^{96}$Zr, we repeat them here with a careful calibration of 
the couplings to transfer channels. 
The basic ingredients of the model are summarized below.

\subsection{Model of neutron transfer} 

One complication in coupled-channels calculations of transfer reactions 
is the enormous number of channels that exist. The calculations can be
simplified by adopting the rotating frame approximation \cite{back} which 
is commonly used in coupled-channels calculations of fusion reactions.
The number of transfer channels is reduced further by lumping them 
together to only one effective channel for each mass partition.
The basic calculation includes zero-, one-, two- and the three-neutron 
transfer channels. Without any influence of inelastic excitations,
it consists of 4 channels and is denoted the Ch-4 calculation.  

The effective form factor for one-neutron transfer is first constructed
from the transfer of the fully occupied $d_{5/2}$ and $s_{1/2}$ states 
in $^{96}$Zr to the unoccupied $f_{7/2}$ state in $^{41}$Ca as described 
in Refs. \cite{landow} using the so-called Quesada form factors 
\cite{quesada}. This effective form factor is denoted $f_{1n}^{\rm eff}(r)$.
The coupling $\langle 1n | V | 0n\rangle$ of the zero- and the one-neutron
transfer channels is assumed to be proportional to $f_{1n}^{\rm eff}(r)$,
\begin{equation}
\langle 1n | V | 0n\rangle = F_{1n} \ f_{1n}^{\rm eff}(r),
\label{V1n}
\end{equation}
where the strength $F_{1n}$ is adjusted so that the one-neutron 
transfer data are reproduced. The reason for this calibration is
that it is very difficult to make a good absolute prediction of 
the one-neutron transfer cross section.

The couplings between the successive one-neutron transfer channels are 
also constructed by simple scaling of the form factor 
$f_{1n}^{\rm eff}(r)$. 
The scaling factors that are used are motivated by the systematics of
transfer reactions that was observed in Ref. \cite{rehm90}.
The basic observation was that the Q value distribution for transfer 
reactions is a Gaussian that is centered at the optimum Q value, which 
is of the order of +1 MeV. The distribution has a maximum cutoff which
is the Q value for the ground state to ground state transition.

The effective ground state Q value for the first one-neutron transfer is 
0.61 MeV \cite{esbcazr}. This means that only about half of the Gaussian 
Q value distribution is accessible because the optimum Q value is close
to +1 MeV. The ground state Q value for two-neutron transfer is $Q_{2n}$ 
= +5.52 MeV \cite{esbcazr}. This implies that the  
full Gaussian Q value distribution is accessible to the second one-neutron 
transfer. The coupling between the one-neutron and the two-neutron transfer 
channels is therefore estimated by
\begin{equation}
\langle 2n | V | 1n\rangle = \sqrt{2} \ F_{1n} \ f_{1n}^{\rm eff}(r).
\label{V2n}
\end{equation}
All of the two-neutron transfer channels are lumped together in the 
coupled-channels calculations into one effective channel and the Q 
value of this channel is set to +1 MeV.

The coupling between the two- and three-neutron transfer channels is set to 
\begin{equation}
\langle 3n | V | 2n\rangle = \sqrt{3/2} \ F_{1n} \ f_{1n}^{\rm eff}(r).
\label{V3n}
\end{equation}
The Q value for the ground state to ground state three-neutron transfer 
is also large and positive ($Q_{3n}$ = +5.24 MeV), and one would therefore 
expect a scaling factor between 1 and the $\sqrt{2}$ in Eq. \ref{V3n}. 
The factor was set to the $\sqrt{3/2}$ in Refs. \cite{landow,esbnisn} and 
that value is also adopted here. The effective Q value for the three-neutron 
transfer is set to +1 MeV in the coupled-channels calculations. 

\begin{figure}
\includegraphics[width = 8cm]{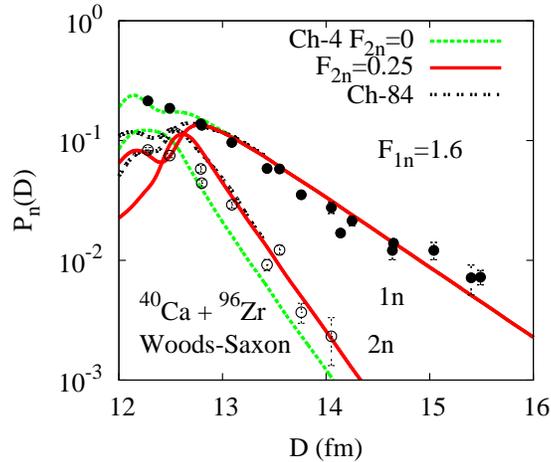}
\caption{ 
Transfer probabilities for the one- and two-neutron transfer
as functions of the distance of closest approach, $D$. 
The data are from Ref. \cite{corradi}. 
All calculations use the one-neutron transfer strength $F_{1n}$ = 1.6.
The successive one-neutron transfer calculation (Ch-4 with $F_{2n}$=0) 
reproduces the one-neutron transfer data at large values of $D$ but it 
does not account for the two-neutron data. The solid curves include
a direct pair transfer with strength $F_{2n}$ = 0.25 fm; it explains 
the two-neutron data quite well at large distances.
The results of Ch-84 calculations are also shown.} 
\label{4096pxtr4}
\end{figure}

The model described above is calibrated so that the measured
one-neutron transfer probabilities of Ref. \cite{corradi} are 
reproduced for large values of the minimum distance $D$ between
the reacting nuclei.
The strength that is required to reproduce the 
data  in Ch-4 calculations is $F_{1n}$ = 1.6 and the results are 
shown in Fig. \ref{4096pxtr4}. It is seen that the two-neutron transfer 
probabilities are under-predicted by the successive Ch-4 calculation 
(the green dashed curve).

In order to reproduce the two-neutron data shown in Fig. \ref{4096pxtr4},
we supplemented the transfer couplings described above with the simple 
pair-transfer coupling originally introduced 
by Dasso and Pollarolo \cite{dasso85},
\begin{equation}
\langle 2n| V | 0n\rangle = - F_{2n} \frac{dU(r)}{dr}, 
\label{V0n2n}
\end{equation}
where $U(r)$ is the nuclear potential. In calculations with up to 
three-neutron transfers we use the same expression, Eq. (\ref{V0n2n}), 
for the coupling between the one-neutron and three-neutron transfer 
channels.  The measured two-neutron transfer probabilities shown in 
Fig. \ref{4096pxtr4} are reproduced quite well at large values of 
$D$ by choosing the pair-transfer strength $F_{2n}$ = 0.25 fm. 
The result is shown by the solid (red) curves.
It is seen that the calculated one-neutron transfer probabilities are not 
much affected by the pair-transfer coupling, except for $D\leq$ 13 fm.

\subsection{Combined effect of excitations and transfer}

The basic assumption of the model developed in Refs. \cite{landow,esbnisn} 
is that inelastic excitations and the neutron transfer are independent 
degrees of freedom. In order to simplify the model it is therefore 
assumed that the excitation spectrum is the same in all of the mass 
partitions that are considered. This implies that if we use the 28 
excitation channels mentioned earlier and combine them with up to 
three-neutron transfers, the full calculation will have 4*28 = 112
channels (Ch-112). If we only include up to two-nucleon transfers, 
there will be 3*28 = 84 channels (Ch-84).   

The Ch-84 calculations that were performed in Ref. \cite{esbcazr} are
repeated here using the one- and two-neutron transfer strengths, 
$F_{1n}$ = 1.6 and $F_{2n}$ = 0.25 fm, that were calibrated in Ch-4
calculations to reproduce the measured one- and two-neutron transfer 
probabilities.
The results of the Ch-84 calculations are shown by the black dashed 
curves in Fig. \ref{4096pxtr4}. It is seen that the transfer probabilities 
are insensitive to the excitations at minimum distances larger than 13 fm
because they are essentially identical to the results of the Ch-4
calculation shown by the solid red curves. The results are different
at smaller minimum distances where fusion can occur and where 
coupled-channels effects are large.  

The fusion cross sections that are obtained in the new Ch-84 calculations
described above are shown by the solid red curves in 
Fig. \ref{4096ffwsch28z}(a) and (b). It is seen that the Ch-84 calculation
does not reproduce the data at low energies (Fig. \ref{4096ffwsch28z}(a)),
and it is slightly above the data at high energies 
(Fig. \ref{4096ffwsch28z}(b)).
Assuming that the Ch-28 model of multiphonon excitations is realistic, 
it appears that the additional couplings to the one- and two-neutron 
transfer channels cannot explain the discrepancy with the measured fusion 
cross sections.
A similar conclusion was recently reached by Scamps and Hagino \cite{scamps}
who also calibrated their transfer couplings to reproduce the neutron 
transfer data shown in Fig. \ref{4096pxtr4} but underestimated the fusion 
cross sections at subbarier energies (see Fig. 11 of Ref. \cite{scamps}.)
We must therefore seek a different explanation for the discrepancy between 
theory and experiment.
Sargsyan et al. suggested that the deformation of the reacting nuclei after 
the two-neutron transfer could explain the data \cite{sargsyan}.
However, they did not test the consistency of their model by comparing
their calculations to the transfer data.
We present in the next section what we believe to be a natural and 
consistent explanation, namely, that one should also consider the effect 
of couplings to one- and two-proton transfer channels.

\subsection{Adjusting the transfer strength.}

\begin{figure}
\includegraphics[width = 7cm]{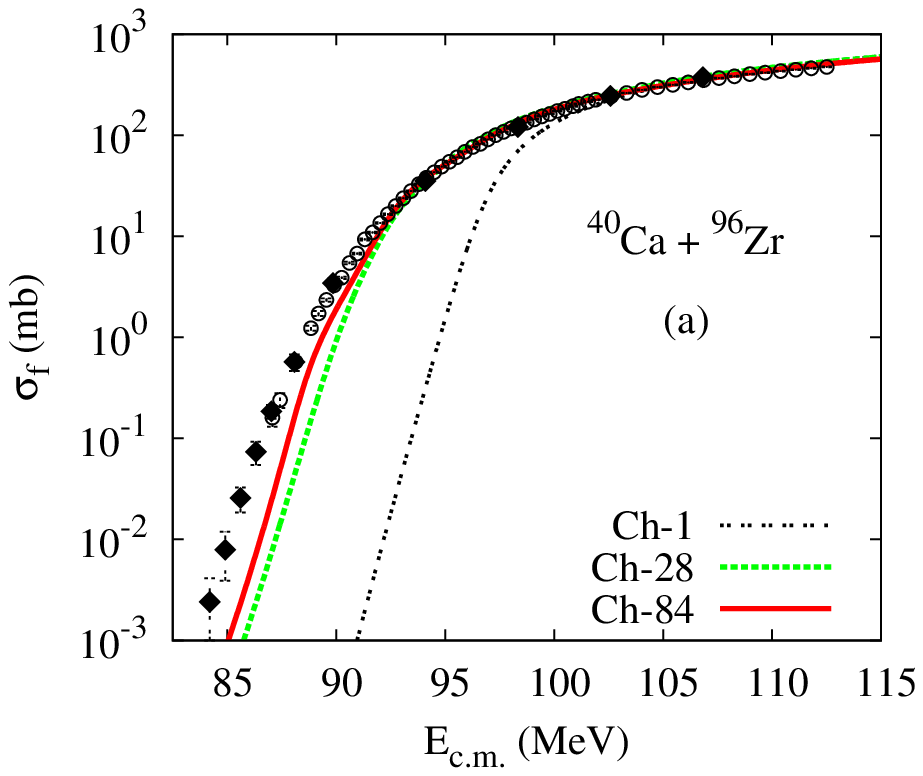}
\includegraphics[width = 7cm]{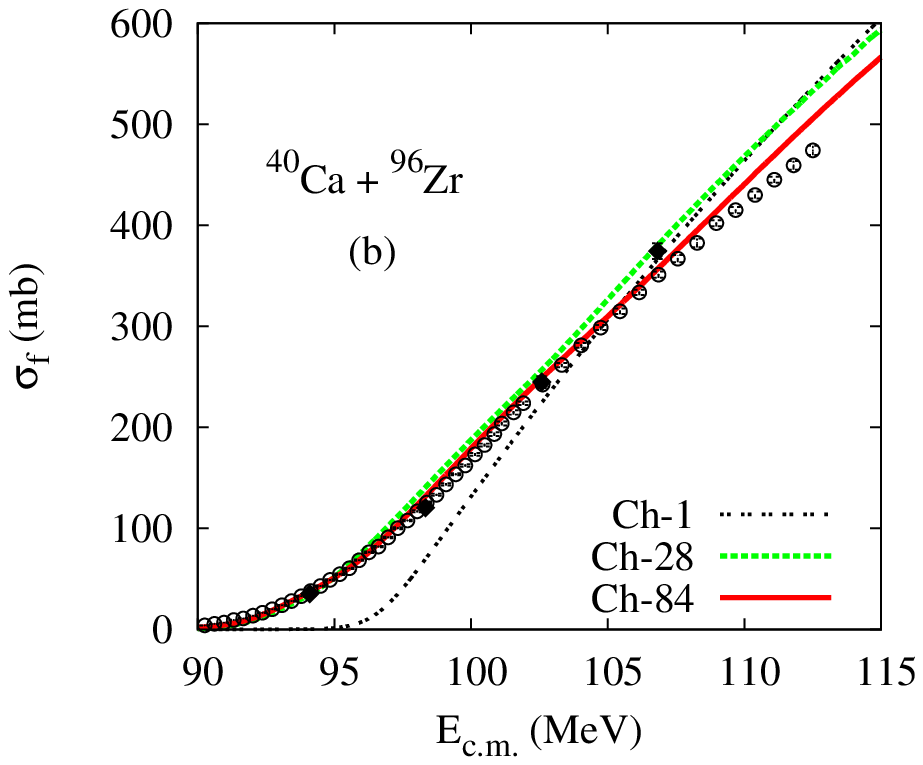}
\caption{
Fusion cross sections for $^{40}$Ca+$^{96}$Zr \cite{timmer,stef4096}  
are compared to coupled-channels calculations that are based on the same 
WS potential as used in Fig. \ref{40zrffws}. 
The Ch-1 calculation is the no-coupling limit. The Ch-28 calculation 
includes couplings of up to three-phonon excitations.
The Ch-84 calculation (solid red curves) uses the transfer strengths 
$F_{1n}$ = 1.6 and $F_{2n}$ = 0.25 fm that were determined 
in Fig. \ref{4096pxtr4}.}
\label{4096ffwsch28z}
\end{figure}

The discrepancy in Figs.  \ref{4096ffwsch28z}(a) and \ref{4096ffwsch28z}(b)
between the Ch-84 calculation and the data can be reduced by increasing the 
transfer strengths. This is clear because a stronger transfer coupling will 
enhance the fusion cross sections at low energies, and it will reduce it 
at high energies. Both features are evidently needed according to the solid 
red curves in Figs. \ref{4096ffwsch28z}(a) and \ref{4096ffwsch28z}(b).
One motivation for increasing the transfer strengths is that the effective 
ground state Q values for the one- and two-proton transfers are positive
according to Table III of Ref. \cite{esbcazr} and couplings to these 
reaction channels could therefore have a significant influence on fusion.
Another motivation is that the fusion data for $^{40}$Ca+$^{96}$Zr and 
other heavy-ion systems were reproduced quite successfully by Pollarolo 
and Winther in applications of their semiclassical method \cite{pol2000}. 
The method includes the combined effects of surface excitations and nucleon 
transfer reactions, and it is likely that the success of the applications 
relied on the inclusion of both neutron and proton transfer.

\begin{figure}
\includegraphics[width = 7cm]{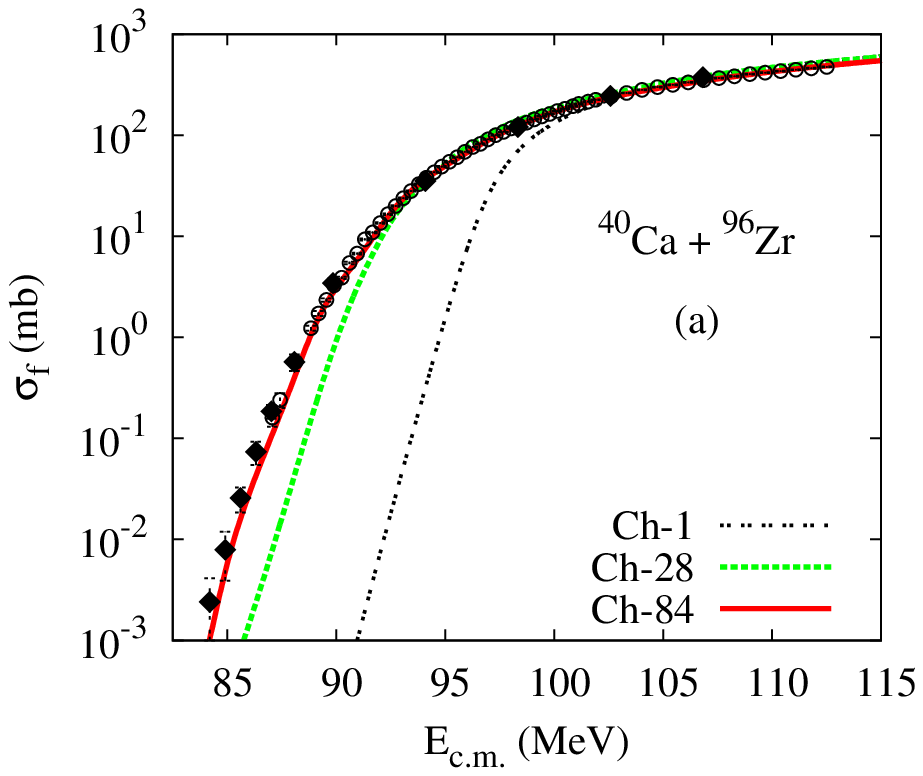}
\includegraphics[width = 7cm]{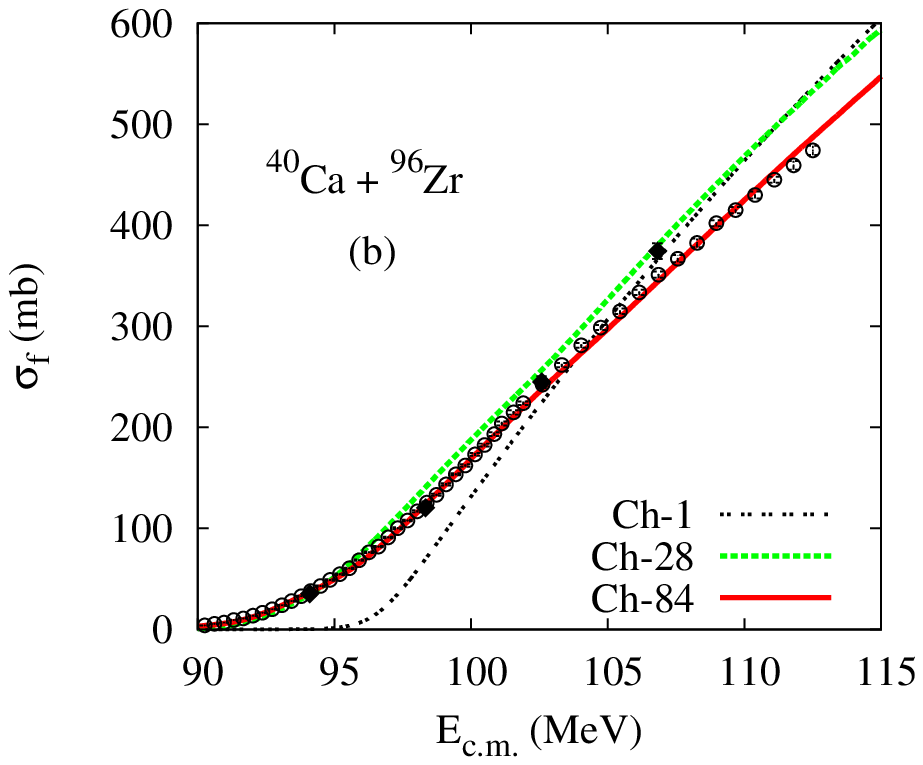}
\caption{ 
Fusion cross sections for $^{40}$Ca+$^{96}$Zr \cite{timmer,stef4096}
are compared to the revised Ch-84 calculations where the transfer couplings
have been multiplied by a $\sqrt{2}$. The results of the Ch-1 and Ch-28 
calculations are the same as those shown in Fig. \ref{4096ffwsch28z}.}
\label{4096ffwsch84x}
\end{figure}

We show in Figs. \ref{4096ffwsch84x}(a) and \ref{4096ffwsch84x}(b) a revised 
Ch-84 calculation in which the transfer coupling strengths were multiplied 
by a factor of $\sqrt{2}$, so that $F_{1n}$ = 2.25 and $F_{2n}$ = 0.355 fm.
This simple scaling is a crude way of simulating the combined effect of 
couplings to neutron and proton transfer channels.
It is seen that the revised Ch-84 calculation reproduce the data very well, 
both at low and at high energies. In fact, the average $\chi^2$ is only 0.87,
assuming a systematic error of 7\%. The fit is much better than obtained 
in Fig.  \ref{40zrffws} with the old Ch-84 calculation of Ref. \cite{esbcazr}, 
and with the new calibrated Ch-84 calculation that is shown in Figs. 
\ref{4096ffwsch28z}(a) and (b). 

The average $\chi^2$ is shown in Table I for each of the three Ch-84 
calculations discussed so far. The energy shift $\Delta E$ that optimizes 
the fit of each calculation is also shown, together with the optimum $\chi^2$.
The non-zero values of the energy shifts $\Delta E$ reflect that the WS 
potential has not been adjusted in each case to minimize the $\chi^2$.
For example, the optimum fit of the revised Ch-84 calculation is achieved 
by applying the energy shift $\Delta E$ = -0.15 MeV to the calculated cross 
section. The shift is equivalent to increasing the radius of the WS well 
by only 0.02 fm.


It is very interesting that we were not able to reproduce the
fusion data in the old analysis of Ref. \cite{esbcazr}, where the assumed
one-neutron transfer coupling was weak ($F_{1n}$=1) and the
two-neutron pair-transfer coupling was adjusted  freely
($F_{2n}$=0.5 fm). In contrast, the new calculations shown in
Figs. 4 and 5 reproduce the data surprisingly well. They use a much
stronger one-nucleon transfer strength ($F_{1n}$=2.25) and a weaker
pair-transfer strength ($F_{2n}$=0.355 fm).
These results demonstrate that the calculated fusion cross sections
are sensitive not only to the pair-transfer coupling but also to the
successive one-nucleon transfer mechanism.

While the couplings to one-nucleon transfer reactions can be calibrated
or tested against transfer data, as it was done in Fig. 2 for the neutron
transfer, the couplings to the successive transfers described by Eqs.
(\ref{V2n}) and (\ref{V3n}) are uncertain or model dependent.
This introduces some uncertainty in the strength of the direct pair-transfer,
which in this work is described by Eq. (\ref{V0n2n}) and is calibrated so
that the combined effect of the successive transfer and the direct
pair-transfer reproduces the measured two-neutron transfer data.
This uncertainty was also discussed in Ref. \cite{scamps}
and needs to be resolved in the future.

Although the revised Ch-84 calculation shown in Fig. \ref{4096ffwsch84x}
is in remarkably good agreement with the data, it is useful to study the 
sensitivity to the multiphonon excitations and to the number of transfer 
channels because the parameters for these reaction channels are uncertain. 
The good agreement with the fusion data could therefore be accidental. 

\section{Dependence on multiphonon excitations and transfer}

It is of interest to study the sensitivity of the calculated fusion
cross sections both to number of multiphonon excitations and to the 
number of nucleon transfers that are considered. Ideally one would 
expect that the most complete calculation in terms of multiphonon
excitations and nucleon transfer channels would provide the best fit
to the fusion data. However, that may not be true in practice because 
of the approximations and model assumptions that have been made. 

It is also important to test the consistency of the calculated fusion
and transfer cross sections and to see if the calculation that provides 
the best fit to the fusion data can also account for the total one- and 
two-nucleon transfer cross sections that have been measured 
\cite{montag2002}. 
All of the calculations that are presented in this section are based on
the WS potential that was described in the beginning of section III.
The calculations that include couplings to transfer channels will be 
based on the revised transfer strengths: $F_{1n}$ = 2.25 and $F_{2n}$ 
= 0.355 fm that were proposed in subsection III.C.

The most obvious way to judge the qualities of the fits to the fusion 
data is to compare the $\chi^2/N$. Another way is to compare the barrier 
distributions obtained from the calculations and the data. It turns
out that the calculated distributions are sensitive to multiphonon 
excitations and to the transfer couplings, and a comparison of the 
experimental barrier distribution may therefore help us identify the 
features that are missing in the calculations.

\subsection{Fusion cross sections}

\begin{figure}
\includegraphics[width = 7cm]{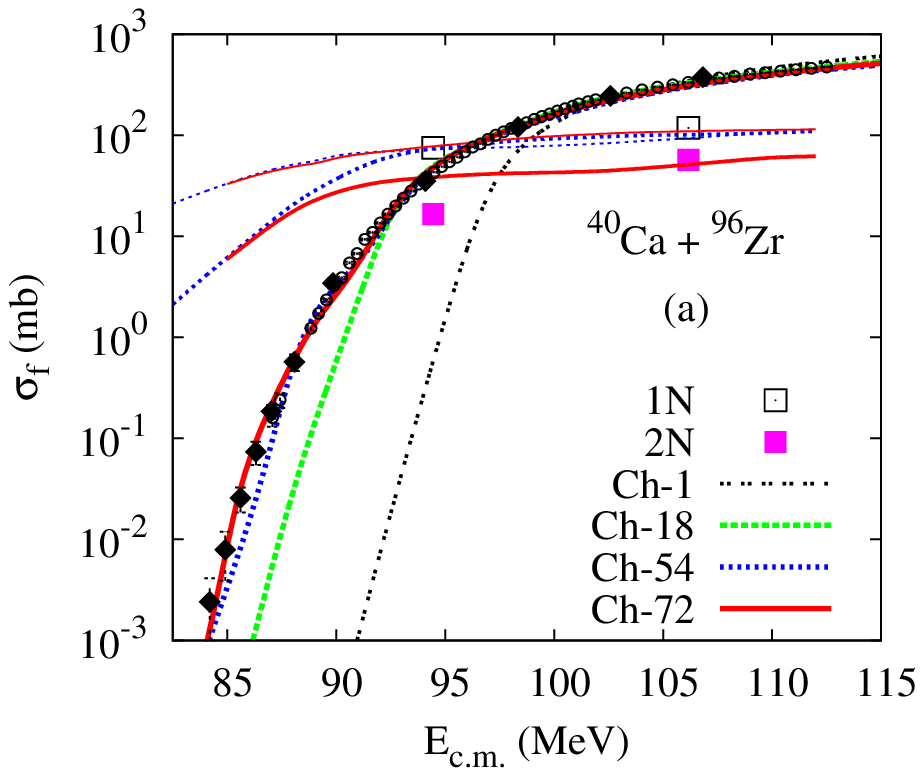}
\includegraphics[width = 7cm]{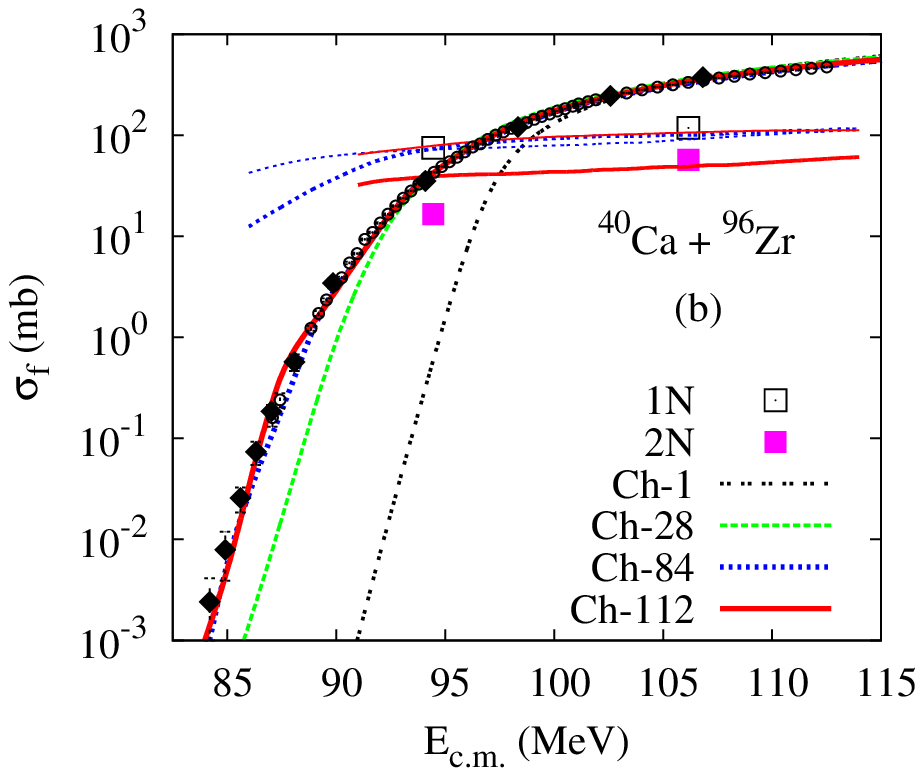}
\caption{ 
Fusion cross sections for $^{40}$Ca+$^{96}$Zr \cite{timmer,stef4096}
are compared to coupled-channels calculations that are based on (a)
Ch-18 and (b) Ch-28 excitation channel calculations.
The Ch-54 and Ch-84 calculations include up to two-nucleon transfers,
and the Ch-72 and Ch-112 include up to three-nucleon transfers.
The transfer strengths were set to $F_{1n}$ = 2.25 and $F_{2n}$ = 0.355 fm.
The measured one-nucleon (1N) and two-nucleon (2N) cross sections 
\cite{montag2002} are compared to the calculations where the thin curves
show the 1N and the thick curves the 2N cross sections.} 
\label{4096ffwsx}
\end{figure}

The calculated fusion cross sections are shown in Fig. \ref{4096ffwsx}.
The coupled-channels calculations shown in Fig. \ref{4096ffwsx}(a) are 
all based on the Ch-18 calculation that includes couplings to one- and 
two-phonon excitations.
The Ch-54 calculation has up to two-nucleon transfers, whereas the Ch-72 
calculation includes up to three-nucleon transfers as explained in 
subsection II.B.
It is seen that the Ch-72 calculation gives the better fit to the fusion 
data, both at low energies and overall in terms of the $\chi^2/N$ that 
is shown in Table I.

The coupled-channels calculations shown in Fig. \ref{4096ffwsx}(b) are 
based on the Ch-28 calculation that includes couplings of up to 
three-phonon excitations. 
It is seen that the Ch-84 and Ch-112 calculations provide better fits to 
the fusion data than do the Ch-54 and Ch-72 calculation that are shown 
in Fig. \ref{4096ffwsx}(a). This implies that multiphonon excitations 
play a very important role in producing a good fit to the data.
The same conclusion was reached in Ref. \cite{esbcazr} for the fusion 
of the other Ca+Zr systems.

The $\chi^2/N$ for the different calculations are compared in Table I. 
The values confirm that the calculated fusion cross sections are sensitive 
to both multiphonon excitations and to the three-nucleon transfer. 
The fact that the Ch-84 calculation and not the Ch-112 calculation gives 
the smallest $\chi^2/N$ is unfortunate and seems to contradict the 
expectation that the most complete calculation should provide the best 
fit to the data. This unfortunate result may be the consequence of the 
approximations and model assumptions we have made. 
For example, the proton and neutron transfer couplings were assumed to 
be similar and their combined effect was estimated by multiplying the 
neutron transfer couplings with a $\sqrt{2}$. 
This estimate may be too crude. In future work it would be desirable also to 
have detailed experimental information about the proton transfer reactions, 
so that one can treat the couplings proton transfer channels explicitly 
and calibrate their strengths to data. 

\subsection{Two-nucleon transfer cross sections}

The one- and two-nucleon transfer cross sections measured at two energies
\cite{montag2002} are shown both in Fig. \ref{4096ffwsx}(a) and (b). 
They are compared to the calculations where the thinner curves indicate 
the results of the Ch-54 and Ch-84 calculations in (a) and (b), and
the thicker curves show the results of Ch-72 and Ch-112 calculations
in (a) and (b), respectively. 
 
It is seen in Fig. \ref{4096ffwsx}(a) that the Ch-72 calculation, which
provides the better fit to the fusion data, also gives the better fit
to the measured one- and two-nucleon transfer cross sections. 
The situation is different in Fig. Fig. \ref{4096ffwsx}(b) where the
Ch-84 calculation gives the better fit to the fusion data whereas the
Ch-112 is in better agreement with the two-nucleon transfer data.
In both cases it is the larger calculation that provides the better 
agreement with the transfer data. This conclusion implies that it is 
important to consider three-nucleon transfer reactions if one wants 
to develop a realistic description of the two-nucleon cross section.

The measured one- and two-nucleon cross sections are compared to the
calculated cross sections in Table II. It is seen that the three-nucleon 
transfer reactions in the Ch-72 and Ch-112 calculations play an important
role in improving the agreement with the data. The influence of
multiphonon excitations is less important. This can be seen by comparing
the results of the Ch-54 and Ch-84 calculations as well as the Ch-72 and 
Ch-112 calculations. Overall, the Ch-112 calculation is in fairly reasonable 
agreement with the measured cross sections, except at the lowest energy
where the calculated two-nucleon transfer cross section is about twice
the measured value.   

\subsection{Barrier distributions}

\begin{figure}
\includegraphics[width = 7cm]{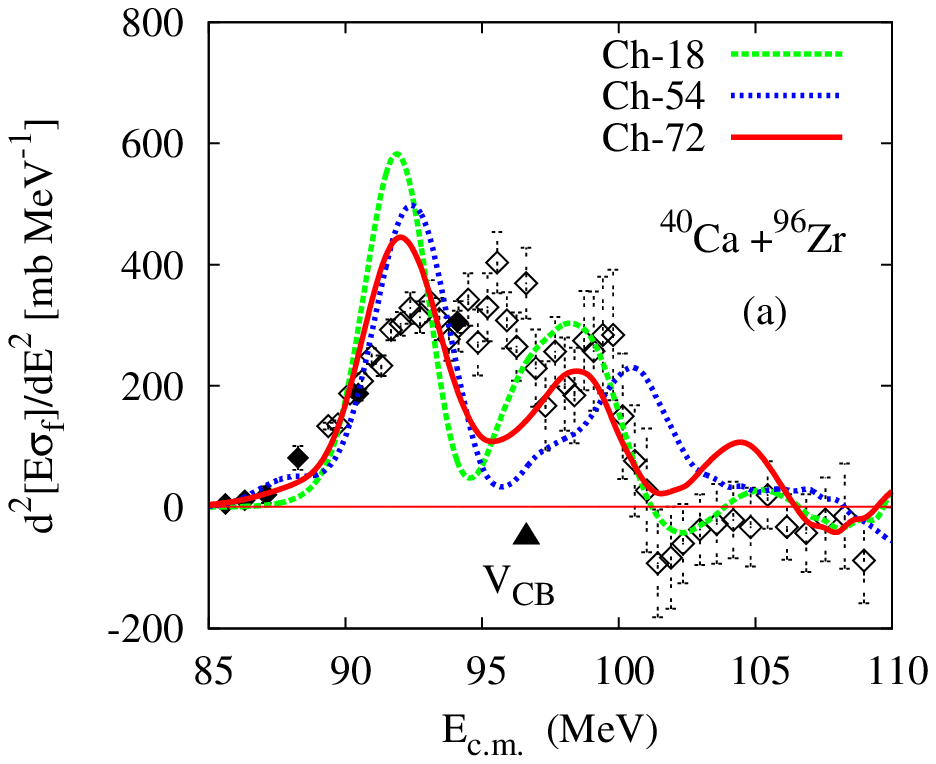}
\includegraphics[width = 7cm]{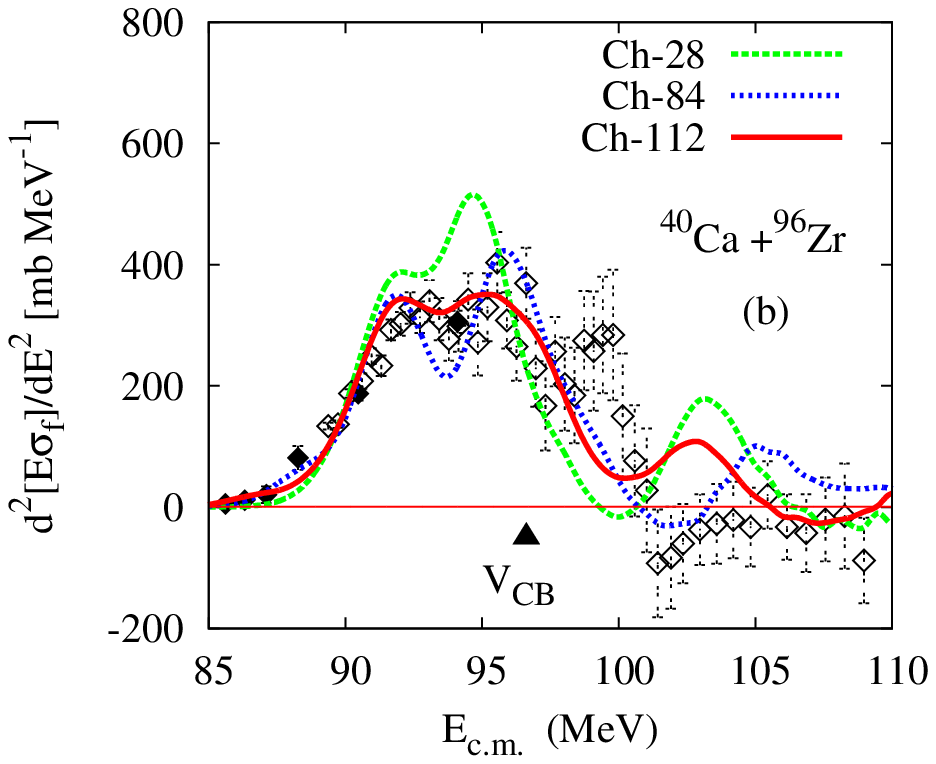}
\caption{ 
Barrier distributions obtained from the fusion cross
sections shown in Fig. \ref{4096ffwsx}.}
\label{4096f2dwsx}
\end{figure}

Another way of illustrating the sensitivity of the calculated fusion cross 
section to multiphonon excitations and transfer reactions is to plot the
derivatives of the cross sections multiplied with the center-of-mass 
energy. The barrier distribution, for example,
is defined as the second derivative \cite{rowley}
\begin{equation}
B(E_{c.m.}) = \frac{d^2(E_{c.m.}\sigma_f)}{dE_{c.m.}^2},
\label{barrier} 
\end{equation}
and it is illustrated in  Figs. \ref{4096f2dwsx}(a) and (b) for the six 
coupled-channels calculations that are shown in Fig. \ref{4096ffwsx}.
The height of the Coulomb barrier in the entrance channel potential is 
indicated by the solid triangle. The calculations show that the couplings 
to multiphonon excitations and transfer channels are both very important
in reproducing the shape of the measured barrier distribution.  
The distributions shown here were calculated using the finite difference
method and an energy step of $\Delta E$ = 2 MeV.

From the comparison of the measured and calculated barrier distributions 
shown in Fig. \ref{4096f2dwsx}(a) and (b) it is clear that the Ch-84 and 
Ch-112 calculations produce the best shapes in comparison to the data. 
This indicates that multiphonon excitations play a very important role
in reproducing the shape of the measured distribution. 
The influence of the couplings to transfer reactions is also important
but the influence of the couplings to the three-nucleon transfer channels 
is modest.
It has the effect of smoothing out certain structures in the barrier 
distribution. This can be seen by comparing the barrier distributions of 
the Ch-84 and Ch-112 calculations. It is seen that the Ch-84 distribution 
has two peaks at energies below the nominal Coulomb barrier, whereas the 
Ch-112 calculation has essentially only one very broad peak. 

We saw earlier that the Ch-84 calculation gives the best $\chi^2$ fit to 
the fusion data, whereas the Ch-112 calculation gives the best agreement 
with the transfer data. From the comparison of the measured and calculated 
barrier distributions it is not so clear which of the two calculations 
gives the best description of the data. A somewhat disturbing feature is 
that the measured distribution has three peaks below 100 MeV, whereas the 
calculations produce at most two peaks. It is not clear at the moment
which reaction mechanism would produce the third peak of the measured
distribution. 

\subsection{$S$ factor for fusion}

\begin{figure}
\includegraphics[width = 10cm]{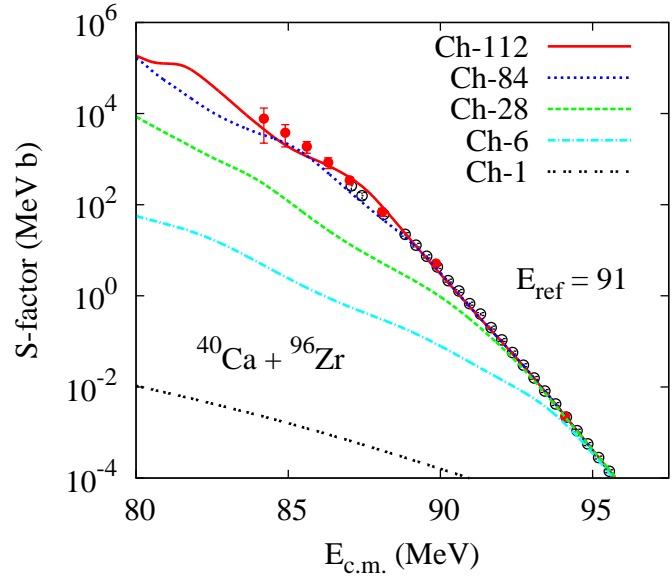}
\caption{
$S$ factors for the fusion cross sections shown in 
Fig. \ref{4096ffwsx}(b).
The reference energy in Eq. (\ref{sfac}) was set to $E_{\rm ref}$ = 91 MeV. 
The second panel shows the results for a deeper WS (Adj WS) potential.}
\label{4096fsfwsx}
\end{figure}

One way to emphasize the behavior of the fusion cross section at low energies 
is to plot the $S$ factor for fusion. It is here defined with respect to a 
reference energy $E_{\rm ref}$ as follows,
\begin{equation}
 S(E_{c.m.}) = E_{c.m.} \sigma_f \ \exp(2\pi[\eta(E_{c.m.}) - \eta(E_{\rm ref})]),
\label{sfac}
\end{equation}
where $\eta(E)$ is the Sommerfeld parameter.  The results based on Ch-28 
calculations are shown in Fig. \ref{4096fsfwsx}.  It is seen that some of 
the calculated $S$ factors exhibit oscillations at the lowest energies. 
The oscillations are sensitive to the depth of the pocket in the entrance 
channel potential and their amplitude can be reduced by choosing a deeper 
pocket. The fact that the $S$ factors obtained from the data do not show
any sign of an oscillation at the lowest energies may indicate that the
pocket in the entrance channel potential is fairly deep.

The most impressive feature of Fig. \ref{4096fsfwsx} is the enormous 
enhancement of the calculated $S$ factors with increasing number of 
channels when compared to the Ch-1 no-coupling calculation.
Another interesting feature is that the data can be reproduced fairly 
well by calculations that use a standard WS potential with a diffuseness
of $a$ = 0.673 fm. There is therefore not any sign of a fusion hindrance 
at the lowest energies, at variance with what has been observed in the
fusion of other heavy-ion systems \cite{back}. The classic example 
is the fusion of $^{60}$Ni+$^{89}$Y \cite{niy} where the data are strongly 
suppressed at low energies compared to coupled-channels calculations that 
are based on a standard WS potential. 
The hindrance has in some systems been so strong that the $S$ factor 
developed a maximum. This is clearly not the case in Fig. \ref{4096fsfwsx}.  

The lack of hindrance in the $^{40}$Ca+$^{96}$Zr fusion data should be 
seen in contrast, for example, to the analysis of the $^{48}$Ca+$^{96}$Zr 
fusion data \cite{stefca48zr} which showed a clear sign of a hindrance at 
the lowest energies \cite{ej4896}. 
The lack of hindrance in the fusion of $^{40}$Ca+$^{96}$Zr correlates 
with the absence of Pauli blocking in transfer reactions near the optimum 
Q value, which is a consequence of the positive Q values for transfer.
The hindrance in the fusion of $^{48}$Ca+$^{96}$Zr, on the other hand, 
correlates with negative transfer Q values and therefore with a Pauli 
blocking of transfer reactions near the optimum Q value.

\section{Conclusions}

In this work we applied and tested a model of heavy-ion fusion and 
transfer reactions that is based on the coupled-channels approach.
The basic assumption is that excitations and nucleon tranfers are 
independent degrees of freedom.  In the application of the model 
it is assumed that the excitation spectrum is the same in all of 
the mass partitions that are considered.

We applied the model to the fusion of $^{40}$Ca+$^{96}$Zr which is
known to be very sensitive to the couplings to multiphonon excitations 
and transfer reactions. 
We first calibrated the transfer couplings so that the measured one- 
and two-neutron transfer probabilities were reproduced at large 
values of the minimum distance between projectile and target.
We showed that the calculation that includes couplings to these transfer 
channels, as well as to multiphonon excitations with up to three-phonon
excitations, cannot explain the fusion data but underestimates them 
substantially at low energies. A similar conclusion was reached in a 
recent work by Scamps and Hagino \cite{scamps}.

In order to explain the fusion data we proposed to increase the strength 
of the transfer couplings. Such an increase is justified because the 
effective Q values for one- and two-proton transfers are positive 
and couplings to these reaction channels should therefore have an 
effect on fusion and enhance it at subbarrier energies.
We assumed for simplicity that the neutron and proton transfers have
similar effects on fusion and simulated their combined effect by 
multiplying the neutron transfer couplings with a factor of the $\sqrt{2}$. 
This estimate, combined with the influence of multiphonon excitations, 
turned out to produce a fusion cross section that is in remarkably good 
agreement with the data.  Moreover, the predicted transfer cross sections 
are in fair agreement with the measured one- and two-nucleon transfer 
cross sections.

It is very interesting that we were not able to explain the 
$^{40}$Ca+$^{96}$Zr fusion data in a previous work \cite{esbcazr},
where the pair transfer strength was adjusted freely in similar 
coupled-channels calculations, without constraining the calculations 
by transfer data. The reason this approach failed must be that the 
assumed single-particle transfer strength was too small.

In future work it would be desirable to measure the neutron and the 
proton transfer cross sections in greater detail, as well as the 
cross sections for other reactions with small or positive Q values.
It would, in particular, be useful to generalize the model we have used 
and treat explicitly the neutron and proton transfer channels, as well 
as other reaction channels that could have an influence on fusion. 
Such a generalization looks very promising in view of the present work.
It should be feasible and fairly straightforward.

{\bf Acknowledgments}.
H. E. was supported by the U.S. Department of Energy, Office of Science,
Office of Nuclear Physics, Contract No. DE-AC02-06CH11357.

\begin{table}
\caption{Analysis of the $^{40}$Ca+$^{96}$Zr fusion data.
The type of calculation is listed in the 1st column.
All calculations use the same WS potential with $R$ = 9.599 fm, 
$V_0$ = -73.98 MeV and diffuseness $a$ = 0.673 fm. 
The strengths of the one- and two-neutron transfer form factors
are listed in the 2nd and 3rd column.
The first $\chi^2/N$ includes all data points. The $\Delta E$ is the
energy shift of the calculation that minimizes the $\chi^2/N$ 
to the data, followed by the value of the minimum  $\chi^2/N$.
The analysis includes a systematic error of 7\%.}
\begin{tabular} {|c|c|c|c|c|c|}
\colrule
 Reaction & $F_{1n}$ & $F_{2n}$ (fm) & $\chi^2/N$  & $\Delta E$ (MeV) & $\chi^2/N$ \\
\colrule
 Ch-84 old     & 1.0  & 0.5   & 4.10 & -0.10 & 3.83 \\ 
 Ch-84 calibr.  & 1.6  & 0.25  & 6.70 & -0.35 & 4.05 \\ 
 Ch-84 revised & 2.25 & 0.355 & 0.87 & -0.15 & 0.30 \\ 
\colrule
 Ch-54        & 2.25 & 0.355 & 2.32 & -0.08 & 2.14 \\ 
 Ch-72        & 2.25 & 0.355 & 1.80 & -0.05 & 1.72 \\ 
 Ch-84        & 2.25 & 0.355 & 0.87 & -0.15 & 0.30 \\ 
 Ch-112       & 2.25 & 0.355 & 1.13 & -0.02 & 1.12 \\ 
\colrule
\end{tabular}
\end{table}

\begin{table}
\caption{The one- and two-nucleon transfer cross sections (in mb) 
measured at 94.5 and 106 MeV \cite{montag2002} are compared to the 
results of the revised coupled-channels calculations that use the 
transfer strengths $F_{1n}$ = 2.25 and $F_{2n}$ = 0.355 fm.}
\begin{tabular} {|c|c|c|c|c|c|c|}
\colrule
 Reaction & $E_{c.m.}$ (MeV) & $\sigma_{1N}$ & $\sigma_{2N}$ & $E_{c.m.}$ (MeV) & $\sigma_{1N}$ & $\sigma_{2N}$ \\
\colrule
 Ch-54      & 94.5 & 73 &  75  & 106.16 &   93  & 101  \\ 
 Ch-72      & 94.5 & 77 &  38  & 106.16 &  109  &  51  \\ 
\colrule
 Ch-84      & 94.5 & 72 &  73  & 106.16 &   90  & 100  \\ 
 Ch-112     & 94.5 & 78 &  39  & 106.16 &  106  &  49  \\ 
\colrule
Ref. \cite{montag2002} & 94.5 & 75$\pm$10 & 16.5$\pm$2.1 & 106.16 &  117.9$\pm$8.5 & 56.9$\pm$4.6 \\
\colrule
\end{tabular}
\end{table}

\end{document}